\documentclass[letterpaper, 10 pt, conference]{ieeeconf}  
\pdfoutput=1

\IEEEoverridecommandlockouts                              

\RequirePackage{luatex85}
\usepackage{times}

\usepackage{soul}
\usepackage{url}
\usepackage[draft]{hyperref}
\usepackage[utf8]{inputenc}
\usepackage[T1]{fontenc}
\usepackage[small]{caption}
\usepackage{amsmath}
\usepackage{amssymb}
\usepackage{booktabs}
\usepackage{pgf}
\usepackage{algorithm}
\usepackage[noend]{algpseudocode}
\usepackage{setspace}
\usepackage{siunitx}
\usepackage[
  toc,
  acronym,
  style = long
]{glossaries}
\makeglossaries

\usepackage[style=ieee,	sorting=none,bibencoding=utf8, maxcitenames=1, mincitenames=1, minbibnames=1, maxbibnames=1, backend=biber, citestyle=numeric-comp, isbn=false,doi=false, url=false, natbib=true]{biblatex}

\graphicspath{{./pictures/}{./figures/}{./pics/}}

\usepackage{graphicx}

\usepackage{array}

\usepackage{booktabs}
\usepackage{longtable}

\usepackage{listings}

\usepackage[colorinlistoftodos,prependcaption,textsize=small]{todonotes}

\usepackage{xargs} 

\usepackage{amsthm}
\usepackage{thmtools}
\usepackage{lscape}
\usepackage{pgfplots}
\usepackage[pass]{geometry}

\declaretheoremstyle[
spaceabove=\topsep, spacebelow=\topsep,
headfont=\normalfont\bfseries,
notefont=\bfseries, notebraces={}{},
bodyfont=\normalfont\itshape,
postheadspace=0.5em,
name={\ignorespaces},
numbered=no,
headpunct=:]
{mystyle}

\usepackage{booktabs}
\usepackage{float}
\usepackage{xcolor}

\usepackage{pdflscape}

\usepackage{footnote}

\DeclareTextFontCommand{\tsf}{\tiny\sffamily} 
\newcommand{\TB}[2]{\ensuremath{T^{#1}_{#2}}}

\newcommand{\ASE}[2]{\ensuremath{e^{#1}_{#2}}}

\newcommand{\clearemptydoublepage}{%
  \ifthenelse{\boolean{@twoside}}{\newpage{\pagestyle{empty}\cleardoublepage}}%
  {\clearpage}}

\newcolumntype{L}[1]{>{\raggedright\arraybackslash}p{#1}}

\newcolumntype{C}[1]{>{\centering\arraybackslash}p{#1}}

\newcolumntype{R}[1]{>{\raggedleft\arraybackslash}p{#1}}

\newcommand{\eq}[1]{Eq.~\ref{#1}}

\newcommand{\roundbr}[1]{\left(#1\right)}

\newcommand\inputpgf[2]{{
\let\pgfimageWithoutPath\pgfimage
\renewcommand{\pgfimage}[2][]{\pgfimageWithoutPath[##1]{#1/##2}}
\input{#1/#2}
}}

\begin{document}

\title{Risk-Based Safety Envelopes for Autonomous Vehicles \\ Under Perception Uncertainty}
\author{Julian Bernhard$^{1}$, Patrick Hart$^{1}$, Amit Sahu$^{1}$, Christoph Sch\"oller$^{1}$ and Michell Guzman Cancimance$^{1}$
   \thanks{$^{1}$All authors are with the fortiss GmbH, An-Institut Technische Universit\"{a}t M\"{u}nchen, Munich, Germany. Correspondence: bernhard@fortiss.org}}

\maketitle
\begin{abstract}
Ensuring the safety of autonomous vehicles, given the uncertainty in sensing other road users, is an open problem.
Moreover, separate safety specifications for perception and planning components raise how to assess the overall system safety.
This work provides a probabilistic approach to calculate safety envelopes under perception uncertainty.
The probabilistic envelope definition is based on a risk threshold. It limits the cumulative probability that the actual safety envelope in a fully observable environment is larger than an applied envelope and is solved using iterative worst-case analysis of envelopes.
Our approach extends non-probabilistic envelopes -- in this work, the Responsibility-Sensitive Safety (RSS) -- to handle uncertainties. To evaluate our probabilistic envelope approach, we compare it in a simulated highway merging scenario against several baseline safety architectures.
Our evaluation shows that our model allows adjusting safety and performance based on a chosen risk level and the amount of perception uncertainty.
We conclude with an outline of how to formally argue safety under perception uncertainty using our formulation of envelope violation risk.
\end{abstract}

\section{Introduction} \label{sec:introduction}
\newacronym{uav}{UAV}{unmanned aerial vehicles}
\newacronym{ai}{AI}{artificial intelligence}

Autonomous systems experience great interest and are developing fast. Examples of such systems include self-driving cars and unmanned aerial vehicles.
The architecture of autonomous systems is frequently based on the sense, plan, and act scheme.
Thereby, sensing and acting are subject to perception and execution uncertainties, e.g., inaccurate state estimates of surrounding vehicles, failures in trajectory tracking, or actuation errors due to environmental influences.

One approach that tries to mitigate this issue is the Simplex architecture that encapsulates components and switches between a nominal (high-performance) and a safety (high-assurance) system using a safety envelope \cite{phan_component-based_2017, desai_soter_2019}.
If the high-performance system proposes actions that violate the safety envelope, the high-assurance system will take over.

However, the definition of safety envelopes is non-trivial.
Simplex architectures and safety envelopes dealing with perception uncertainty often use metrics over the output of the perception system, e.g., a neural network object detector, to decide whether the high-assurance system should take over or not~\cite{cofer_run-time_2020}.
The definition of such metric and appropriate take-over thresholds only by considering faults at the perception level is questionable since often no direct conclusion over the actual safety of the system can be drawn.

Various approaches focus on the definition of provably safe envelopes for planning algorithms that are decoupled from the perception task.
These envelopes mostly rely on physical principles and safe distances \cite{shalev-shwartz_formal_2017, yu_risk_2020}, reachable sets \cite{pek_provably_2020}, fail-safe trajectories \cite{pek_computationally_2018} or constrain the risk of violating safety envelopes to target safety under prediction uncertainty \cite{bernhard_risk-constrained_2021}.
However, these approaches can only provide guarantees given perfect perception without uncertainties.

\begin{figure}[t]
	\vspace{0.3cm}
	\def\svgwidth{1.0\columnwidth}
	\centering
\begingroup%
  \makeatletter%
  \providecommand\color[2][]{%
    \errmessage{(Inkscape) Color is used for the text in Inkscape, but the package 'color.sty' is not loaded}%
    \renewcommand\color[2][]{}%
  }%
  \providecommand\transparent[1]{%
    \errmessage{(Inkscape) Transparency is used (non-zero) for the text in Inkscape, but the package 'transparent.sty' is not loaded}%
    \renewcommand\transparent[1]{}%
  }%
  \providecommand\rotatebox[2]{#2}%
  \newcommand*\fsize{\dimexpr\f@size pt\relax}%
  \newcommand*\lineheight[1]{\fontsize{\fsize}{#1\fsize}\selectfont}%
  \ifx\svgwidth\undefined%
    \setlength{\unitlength}{223.08545931bp}%
    \ifx\svgscale\undefined%
      \relax%
    \else%
      \setlength{\unitlength}{\unitlength * \real{\svgscale}}%
    \fi%
  \else%
    \setlength{\unitlength}{\svgwidth}%
  \fi%
  \global\let\svgwidth\undefined%
  \global\let\svgscale\undefined%
  \makeatother%
  \begin{picture}(1,0.50977091)%
    \lineheight{1}%
    \setlength\tabcolsep{0pt}%
    \put(0,0){\includegraphics[width=\unitlength,page=1]{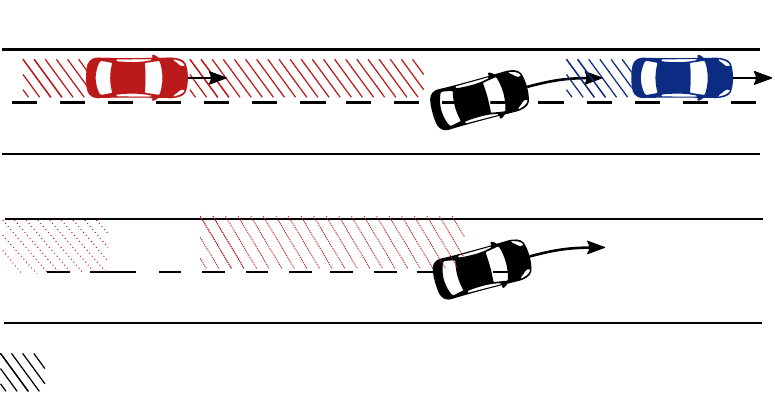}}%
    \put(0.23956648,0.48800957){\color[rgb]{0,0,0}\makebox(0,0)[t]{\lineheight{0}\smash{\begin{tabular}[t]{c}\sffamily\scriptsize Without perception uncertainty\end{tabular}}}}%
    \put(0.06430617,0.01868562){\color[rgb]{0,0,0}\makebox(0,0)[lt]{\lineheight{0}\smash{\begin{tabular}[t]{l}\sffamily\tiny unsafe region\end{tabular}}}}%
    \put(0.21758252,0.25597725){\color[rgb]{0,0,0}\makebox(0,0)[t]{\lineheight{0}\smash{\begin{tabular}[t]{c}\sffamily\scriptsize With perception uncertainty\end{tabular}}}}%
    \put(0,0){\includegraphics[width=\unitlength,page=2]{perception_uncertainty.pdf}}%
    \put(0.318861,0.01567606){\color[rgb]{0,0,0}\makebox(0,0)[lt]{\lineheight{0}\smash{\begin{tabular}[t]{l}\sffamily\tiny unsafe region under uncertainty\end{tabular}}}}%
    \put(0,0){\includegraphics[width=\unitlength,page=3]{perception_uncertainty.pdf}}%
  \end{picture}%
\endgroup%

	\caption{The difficulty of defining safety envelopes in the presence of perception uncertainty.
	When neglecting perception uncertainty, safe regions for an autonomous vehicle can be defined using physical limits of vehicle dynamics (top). However, all measurements are subject to perception uncertainty, making a strict definition of unsafe regions impossible (bottom).
	We address this problem with a probabilistic definition and calculation of safety envelopes.}\vspace{-0.3cm} \label{fig:perception_uncertainty}
\end{figure}

To address these problems, we propose a probabilistic approach to calculate safety envelopes under perception uncertainty.
Our envelope definition is based on a risk threshold, limiting the cumulative probability that the actual safety envelope is larger than the applied safety envelope under perception uncertainty, and solved using iterative worst-case analysis of envelopes.
Specifically, we contribute
\begin{itemize}
	\item a formalization of the probabilistic safety envelope under perception uncertainty,
	\item a method to calculate a safety envelope satisfying a given risk-level by using iterative worst-case combinations and an approximation of cumulative distributions over envelopes,
	\item a statistical, simulative evaluation of the proposed approach in the domain of autonomous driving.
\end{itemize}

The implementation of our probabilistic envelope builds upon the RSS safety specification \cite{shalev-shwartz_formal_2017}.
However, other safety envelope definitions that assume independence between agents can be used with the proposed approach.

\section{Related Work} \label{sec:behavior_modeling}
\newacronym{dnn}{DNN}{deep neural network}
\newacronym{rss}{RSS}{responsibility-sensitivity safety metric}
\newacronym{rtsa}{RTSA}{run-time safety assurance}
\newacronym{brs}{BRS}{backward reachable set}
There are various approaches proposed in the literature to verify the safety of autonomous vehicles. Reachability analysis, invariant sets, probabilistic approaches, formal logic using theorem proving, logical formulas, and more.
This section gives a brief outline of the related work and how our proposed methodology compares to the state-of-the-art.

\subsection{Simplex Architecture}
A popular framework for run-time safety assurance is the Simplex architecture \cite{Sha2001}.
It typically consists of a \emph{high-performance} and a \emph{high-assurance} controller.
The high-performance controller can be any component, such as learning-based ones using neural networks.
The high-assurance controller is developed adhering to standards and with safety in mind.
A \emph{decision logic} then switches between these based on some defined boundary.
In the literature, various ways have been employed for defining these decision boundaries ranging from statically defined safe regions \cite{Desai2019}, over reachability analysis \cite{Pek2020}, to reactive synthesis \cite{Mitsch2016}.
Many Simplex architectures suffer from control boundary switching as the high-performance controller is unaware of being restricted, leading to  deteriorating performances.
The \gls{rss} framework tries to mitigate this issue by not only switching between the high-performance and high-assurance controller but by additionally restricting the accelerations the vehicle can take a-priori \cite{Shalev-Shwartz2017}.
Similarly, SOTER \cite{Desai2019} has each component in the system defined as a \gls{rtsa} module that makes sure that the sub-system operates within the pre-defined bounds.
However, SOTER is only demonstrated in static environments with pre-defined safe regions.
Other variations include using past safe trajectories for recovering safe operational states~\cite{Mehmood2021} or learning the control boundary for increasing overall performance~\cite{Lazarus2020}.

Most approaches discussed do not offer a holistic view on integrating uncertainties into the overall Simplex architecture.
However, we argue that for achieving truly safe autonomous systems in dynamic environments, Simplex architectures have to propagate uncertainties and use these in the reasoning for, e.g., risk-based decision switching.

\subsection{Safety Envelope Definition}
Safety envelopes serves as switching boundary between high-performance/high-assurance controller. Apart from that they providently yield safe behavior by restricting the high-performance function to not violate the safety envelope in the near future. There are variations in the literature on how to define such envelopes.

\textcite{pek_provably_2020} defines a safety envelope in terms of driving corridors and uses these for fail-safe trajectory planning by computing the drivable area of the ego vehicle using forward reachability analysis.

\textcite{leung_infusing_2020} define envelope conditions in terms of a persistent collision-free escape maneuver under worst-case actions by other cars.
They use backward reachability analysis to compute a \gls{brs}.
This \gls{brs} consists of states from where the controller will not be able to prevent the reachability into an undesirable (collision) state within a time horizon $T$.

The envelope definition used in our work is based on the concepts of the \gls{rss} model \cite{Shalev-Shwartz2017}.
\gls{rss} is based on five safety rules: Safe distance, cutting in, right of way, limited visibility, and avoid crashes. The safety envelope is built upon the definition of safe distances (Section~\ref{sec:background}).

Assuming Gaussian uncertainty state estimates, \textcite{Nyberg2021} apply Taylor expansion to obtain an analytical form of the probability density of longitudinal safe distances. As in our work, they measure the probability of violating the true safe distance, and additionally model the severity of envelope violations as coherent risk measure. Yet, they do not provide a safety envelope for their risk formulation. Their planner integrates the risk measure into a single-objective cost criterion. In contrast, we calculate a safety envelope fulfilling a given maximum violation risk constraint which, similar to \gls{rss}, restricts the behavior of a performance-driven planner. 

\textcite{Klaes2021} uses the concept of an uncertainty wrapper \cite{uncertaintyWrapper} in a safety monitor based on \gls{rss}.
They measured safe distance (Eq.~\ref{eq:safe_distance}) more precisely by constraining uncertainty in maximum braking $\beta_{max}$ using dynamic estimation of friction coefficient $\mu$.
The uncertainty wrapper is most similar to the concept that we propose in this paper.
In contrast to basing uncertainty on an additional variable $\mu$, we directly define and use the perception uncertainty.

\section{Background}
\label{sec:background}

This section presents the background of safety envelopes and how these can, e.g., be defined using analytical equations and safe distances to other vehicles.
Consider a single lane scenario with a leading vehicle as an example.
In such a scenario, analytical equations exist for the ego vehicle to maintain a safe distance to other vehicles.
Let $c_r$ be a vehicle that is behind $c_f$ in the longitudinal coordinates.
Considering $\rho$ as response time and assuming acceleration limits on both vehicles as follows: $\beta_{max}$ for maximum braking by $c_f$, $a_{max}$ for maximum acceleration during response time by $c_r$, and $\beta_{min}$ as minimum braking afterwards by $c_r$.
The minimum longitudinal distance that must be maintained by the rear vehicle $c_r$ to not cause any collision is defined in \cite{Shalev-Shwartz2017} as:

\begin{equation} \label{eq:safe_distance}
d_{min} = \left[ v_r\rho + \frac{1}{2} a_{max}\rho^2 + \frac{\left( v_r + \rho a_{max} \right)^2}{2 \beta_{min}} - \frac{v_f^2}{2 \beta_{max}}         \right]
\end{equation}

\begin{figure}[t]
	\def\svgwidth{1.0\columnwidth}
	\centering
	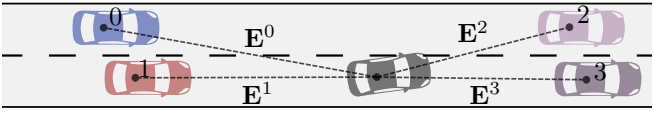
	\caption{Star shaped calculation for the pair-wise safety envelopes $\mathbf{E}$.
	The overall response returns the most restrictive constraints of all pair-wise safety envelopes.
	} \label{fig:safety_envelope}
\end{figure}

The longitudinal and similarly defined lateral safe distances from all other vehicles in a scenario impose constraints on the maximal allowed longitudinal and lateral accelerations.
A star shaped calculation, as shown in Figure~\ref{fig:safety_envelope} computes the pair-wise acceleration limits between ego vehicle $i$ and each other vehicle $j$. The minimum of constraints on the maximum accelerations and decelerations is considered to be the safety envelope of the ego vehicle.

In the remainder of this paper, we assume that such an envelope definition is available, e.g., for autonomous vehicles defined by \gls{rss} \cite{Shalev-Shwartz2017}.
The \gls{rss} returns for a given state $\mathbf{s}_t$ and time span $\tau$ a vector of acceleration limits $\mathbf{E}$ with
\begin{equation}
\mathbf{E} = \operatorname{env}(\mathbf{s}_t, \tau)
\end{equation}
allowing to plan strictly safe decisions for duration $\tau$ by staying within these limits.
In the case of \gls{rss}, the vector consists of longitudinal and lateral maximum and minimum accelerations $\mathbf{E}=(a_\text{min,lat}, a_\text{max,lat}, a_\text{max,lon}, a_\text{max,lon})$. We also define an indicator function $f_\text{safety}: \mathcal{S} \rightarrow  \{0, 1\}$ that yields zero if the state violates the safety envelope.

Given perfect perception, without uncertainty, restricting the dynamics of the autonomous system to stay within $\mathbf{E}$ does not cause any collisions as has been empirically shown in \cite{shalev-shwartz_formal_2017}.
However, in practice, the assumption of perfect perception is unrealistic.
Our method extends non-perception-based envelope definitions such that even in the case of perception uncertainty, the autonomous system is safe up to a pre-defined probability.

\section{Problem Formulation}
\label{sec:problem_formulation}
\newacronym{cdf}{CDF}{cummulative distribution function}
\newacronym{rss}{RSS}{responsibility-sensitive-safety}

We consider a multi-agent environment in which an ego agent $i$, the autonomous vehicle or drone, interacts with $N$ other agents denoted by $j$.
The true state of the environment
\begin{equation}
	\mathbf{s}_t = (\mathbf{s}^0_t, \dots, \mathbf{s}^n_t)
\end{equation}

at time $t$ is not observable.
We assume that the ego agent $i$ can perfectly localize itself and no occlusions occur.
The ego agent then senses its environment and fuses the received sensor data into an observed environment state

\begin{equation}
	\mathbf{\hat{s}}_t = (\mathbf{o}^0_t, \dots, \mathbf{o}^n_t).
\end{equation}

This observed state $\mathbf{\hat{s}}_t$ consists of individual observations for each other agent $\mathbf{o}_t^j = (x^j, y^j, v^j, \theta^j)$ with $j \neq i$, where $(x^j, y^j)$ denotes the agent's position, $ \theta^j$ its orientation and $v^j$ its velocity in direction of orientation.

Since all measurements $\mathbf{o}^j_t \neq \mathbf{o}^i_t$ are subject to noise, the observed environment state differs by a certain extent from the true environment state, i.e.

\begin{equation} \label{eq:observed_state_world_dist}
	\mathbf{\hat{s}}_t = \mathbf{s}_t  + \mathbf{\delta}_t \text{ with } \mathbf{\delta}_t \sim \mathbf{P}_\delta.
\end{equation}

Here we assume that the deviation $\mathbf{\delta}_t$ of $\mathbf{\hat{s}}_t $ from the true environment state $\mathbf{s}_t$ follows some distribution $\mathbf{P}_\delta$.
We further assume that the noise distribution $\mathbf{P}_\delta$ is the same at each timestep $t$.

In modern autonomous systems, it is the goal of the perception component to minimize the noise $\mathbf{\delta}_t$ and, as a result, to be as accurate as possible.
However, zero noise cannot be achieved, especially due to aleatoric uncertainties that are not removable (e.g., motion blur in images). The perception component should estimate $\mathbf{P}_\delta$ and provide the estimated perception uncertainty $\mathbf{\hat{P}}_\delta$ along with the state estimates $\mathbf{o}_t$ to the system's planning component to account for these errors.

The planner then takes this uncertainty estimate into account to define safety margins and generate a safe future trajectory $T$. The trajectory $T$ is an ordered list of agent states. After the trajectory is planned, the system's controller -- which we assume to be ideal -- then generates commands to follow this trajectory exactly.

In the case of full observability of the environment ($\forall t, \mathbf{\delta}_t = 0$), a safety envelope $\operatorname{env}(\mathbf{s}_t, \tau)$ can be used to plan strictly safe trajectories at time $\mathbf{s}_t$ for a timespan $\tau$.
Yet, in the case of added perception noise ($\exists t, \mathbf{\delta}_t \neq 0$), such envelope definitions do not allow for a quantitative safety argumentation.

In this paper, we aim to extend physical safety envelopes to allow for a quantitative safety argumentation in the context of perception uncertainty. We assume an ideal perception that correctly estimates the true measurement noise distribution, i.e. $\mathbf{\hat{P}}_\delta = \mathbf{P}_\delta$.

A safety envelope $\mathbf{E_\text{percept}} = \operatorname{EnvPercept}(\mathbf{\hat{s}}_t)$ considering perception noise can only be defined based on the observed state $\mathbf{\hat{s}}_t$.
From the perspective of the ego vehicle the true state $\mathbf{s_t}$ and envelope $\mathbf{E}$ are \emph{random variables}. Given this, we can define a requirement on $\mathbf{E_\text{percept}}$:
\begin{equation} \label{eq:perception_envelope}
\text{Pr}_{\delta_t \sim P_{\delta}}[\mathbf{E} < \mathbf{E_\text{percept}} ] \overset{!}{\leq} \beta_\text{risk}.
\end{equation}
The probability that the probabilistic envelope $\mathbf{E}_\text{percept}$ is larger, i.e., less restrictive than the true envelope $\mathbf{E}$ must be lower than the required risk level $\beta_\text{risk}$\footnote{In this work, we use the term risk to model the probability of violating a safety envelope.
Without loss of generality, the severity of envelope violations, e.g., modeled in \cite{Nyberg2021}, can be integrated into our problem formulation.} when observed states are distributed according to Eq.~\ref{eq:observed_state_world_dist}.
The greater equal comparison is interpreted component-wise and interpreted such that deceleration and acceleration signs are handled in a safety consistent way.
Note that it is not feasible to resolve the equation analytically because safety envelope definitions such as RSS are highly non-linear.
Also, we must find a way to safely approximate the random envelope variable $\mathbf{E}$ to solve this requirement.

In this work, we design an algorithm $\operatorname{EnvPercept}(\mathbf{\hat{s}}_t)$, which is based on a given safety envelope definition $\operatorname{env}(\mathbf{s}_t, \tau)$. It returns a probabilistic safety envelope satisfying Eq.~\ref{eq:perception_envelope}.
This work assumes no time delays due to perception and planning processing and a guaranteed re-planning phase every $\tau$ seconds.

\section{Method}
\label{sec:method}

\begin{figure*}[!t]
	\def\svgwidth{1.0\textwidth}
	\centering
	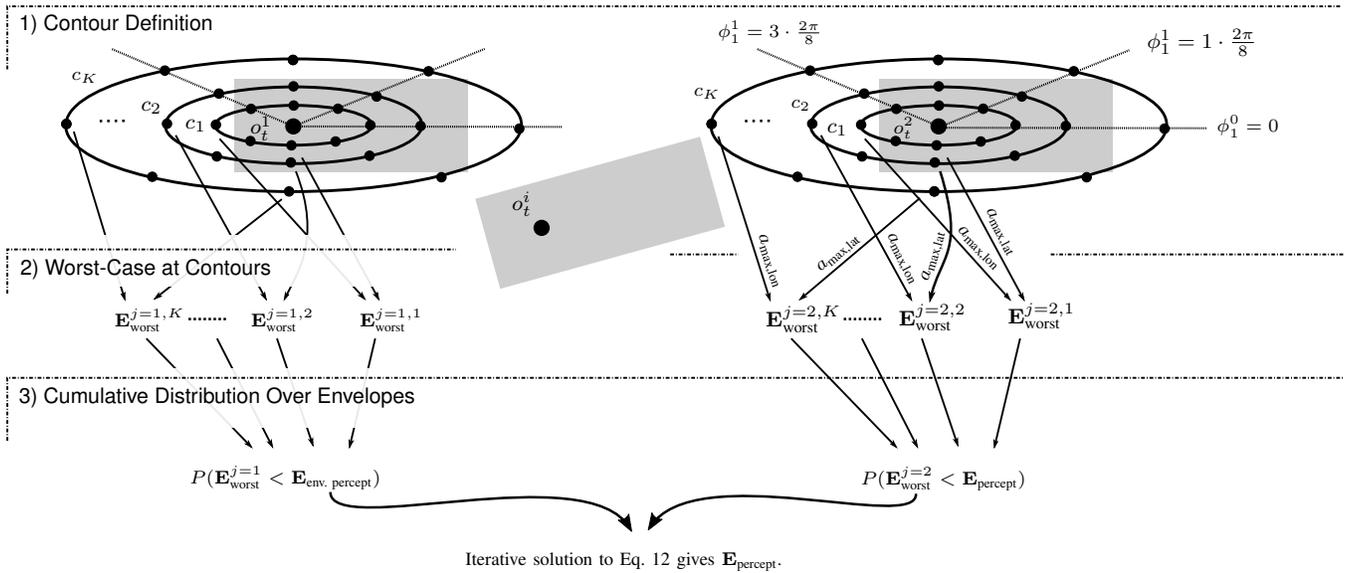
	\caption{Overview of the probabilistic envelope calculation for two other agents and the ego agent at observed state $\mathbf{\hat{s}}_t=(o_t^i, o_t^{j=1}, o_t^{j=2})$.
	1) State-confidence contours $c_k,\, k\in \{1,\ldots, K\}$ are located at observed states $o_t^j$ of each other agent.
	2) Worst-case envelopes $\mathbf{E}_\text{worst}^{j,k}$ for each contour are obtained by evaluating the envelope function $\operatorname{env}(\cdot)$ at the contours using fixed angular sampling points $\phi_k^z$ and a component-wise comparison of the acceleration restrictions.
	3) Worst-case envelopes at contours $\mathbf{E}_\text{worst}^{j,k}$ define a random envelope variable $\mathbf{E}_\text{worst}^{j}$ corresponding to each agent.
	$\mathbf{E}_\text{percept}$ is then obtained in an iterative procedure solving Eq.~\ref{eq:cdf_envelopes}.}	\label{fig:method_overview}
\end{figure*}

An overview of our method is given in Figure~\ref{fig:method_overview} that consists of the following three steps:
\begin{enumerate}
	\item \textbf{Definition of State-Confidence Contours:} Using cumulative distribution functions, we obtain confidence regions for each agent's state deviations.
	\item \textbf{Worst-case Envelopes at Contours:} For the contour of each state-confidence region, a fixed set of state samples is evaluated to generate the envelope $\operatorname{env}(\cdot)$. The worst-case envelopes of each agent define a random envelope variable given the probability that the state deviations is in a certain region. 
	The worst-case envelope is returned using a component-wise acceleration comparison.
	\item \textbf{Cumulative Distribution Over Envelopes:} The minimum over random envelopes of all agents (solved for each acceleration component individually) defines a combined random envelope variable approximating the true envelope variable $\mathbf{E}$. Eq.~\ref{eq:perception_envelope} is then solved iteratively by accumulating envelope probabilities to obtain $\mathbf{E_\text{percept}}$. 
\end{enumerate}

The above-outlined steps enable quantifiable safe behaviors for autonomous systems in uncertain environments.
We describe these steps in detail in the following.

\subsection{Definition of State-Confidence Contours} \label{sec:contour_lines}
Our system receives the perceived state $\mathbf{\hat{s}}_t$ with observations $(\mathbf{o}^0_t,\dots,\mathbf{o}^n_t)$ at timestep $t$ from its perception component.
Additionally, our method receives the perception uncertainty $\mathbf{\hat{P}}_\delta$.
Instead of a joint distribution, we assume independence and model $\mathbf{\hat{P}}_\delta$ as a set of normal distributions $(\hat{\mathcal{N}}^0_t,...,\hat{\mathcal{N}}^n_t)$ for every perceived agent.
All $\hat{\mathcal{N}}^j_t$ are centered at their corresponding observation $\mathbf{o}^j_t$ and have the same uncertainty covariance $\mathbf{\Sigma}$\footnote{to simplify notation, the approach straightforwardly extends to differing covariances}.
Given these definitions, we define uncertainty contours $c_k$ at fixed iso-probability levels $p_k,\, k\in\{1,\ldots,K\}$ around each agents' observed state $\mathbf{o}^j_t$.
A confidence region of a multivariate normal distribution $\mathcal{N}^j_t$ is defined by its cumulative distribution function and has the shape of an ellipsoid.
Then, assuming $\mathcal{N}^j_t$ is axis-aligned, the contour $c_k$ that contains the vehicle's true state $\mathbf{s}^j_t$ with probability $p_k$ is defined as the hyper-ellipsoid level-set
\begin{equation}
\label{eq:normal-ellipse}
\roundbr{\frac{\Delta x^j}{\sigma_x}}^2 + \roundbr{\frac{\Delta y^j}{\sigma_y}}^2 + \roundbr{\frac{\Delta v^j}{\sigma_{v}}}^2 + \roundbr{\frac{\Delta \theta^j}{\sigma_{\theta}}}^2 = \mathcal{Q}_4(p_k),
\end{equation}
where $\sigma$ are the axis-wise standard deviations, $\Delta$ represent the deviations to the observed state $\mathbf{o}^j_t$ and $\mathcal{Q}_4(p_k)$ is the the quantile function of fourth degree Chi-square distribution $\mathcal{X}^2_4$, i.e., the inverse of its cummulative distribution function.

In most cases, the distribution $\hat{\mathcal{N}}^j_t$ will not be axis-aligned.
To be still able to obtain the ellipsoid confidence region, we first compute the eigendecomposition of $\Sigma$ and then use the four resulting eigenvalues that represent the axis-aligned variances. By computing the square root of these eigenvalues, we obtain the necessary standard deviations for \eq{eq:normal-ellipse}.

\subsection{Worst-Case Envelopes at Contours}
As most safety envelopes are highly non-linear there is no closed form solution available for calculating the envelopes for all continuous states at the contours.
Therefore, we sample states on the contour $c_k$ for a specified iso-probability $p_k$ at fixed angular variations.
Note that the state deviations on the hyper-ellipsoid of dimension $N=4$ defined in Eq.~\ref{eq:normal-ellipse} can be obtained by iterating through $N-1=3$ angles using

\begin{equation}
 \delta_k(\phi_1, \phi_2, \phi_3) = \begin{bmatrix}
	x \\
	y \\
	\theta \\
	v
 \end{bmatrix} = \begin{bmatrix}
	r_0\cos(\phi_1) \\
	r_1\sin(\phi_1)\cos(\phi_2) \\
	r_2\sin(\phi_1)\sin(\phi_2)\cos(\phi_3) \\
	r_3\sin(\phi_1)\sin(\phi_2)\sin(\phi_3)
 \end{bmatrix},
\end{equation}
with axis-wise radii
\begin{equation}
	r_i = \sqrt{\mathcal{Q}_4(p_k)\lambda_i},
\end{equation}
and with $\lambda_i$ being the eigenvalues of the covariance matrix $\mathbf{\Sigma}$ after eigendecomposition.
All sampled states have to be evaluated, as the true state $\mathbf{s}_t$ is within the given confidence hyper-ellipsoid with $p_k$ probability.
Therefore, we employ a worst-case consideration for each ellipsoid contour. This means we compute the envelope for each sample on the contour for a specific agent. Then we find the combined envelope with limits such that no envelope of the contour samples would be violated.

Given these considerations, we obtain a single worst-case envelope for each contour $c_k$ for all agents $i$.
The contour is evaluated at discrete angles $\phi_l^z = z\cdot2\pi/N_\phi,\, z\in\{0, N_\phi-1\},\, l\in\{1,2,3\}$ with $N_\phi$ being the number of evaluated angles for each dimension.
At each set of discrete angles, envelopes corresponding to agent $j$ are obtained
\begin{equation}
	\mathbf{E}^{j,k}(p, q, r) = \operatorname{env}(\mathbf{o}_t^j + \mathbf{\delta}_k(\phi_1^p, \phi_2^q, \phi_3^r), \tau)
\end{equation}
and a component-wise comparison of acceleration limits gives the worst-case envelope $\mathbf{E}_\text{worst}^{j,k}$ corresponding to agent $j$ for contour $c_k$.
For instance, in the case of the RSS, we compare $a_\text{min,lat}(p, q, r), a_\text{max,lat}(p, q, r), a_\text{max,lon}(p, q, r)$ and $a_\text{max,lon}(p, q, r)$ for all combinations of $p,q,r$ and use the most restricting limits to define the worst-case envelope $\mathbf{E}^{j,k}_{worst}$.

\subsection{Cumulative Distribution Over Envelopes} \label{sec:cdf_envelopes}
In the probabilistic setting, the star-shaped calculation over deterministic envelopes (cf.~\ref{sec:background}) transfers to a minimum operation over random envelope variables of each agent yielding a combined random envelope variable. The probabilistic envelope is calculated such that the cumulative probability of the combined envelope variable is below $\beta_\text{risk}$. These steps are outlined in detail in the following.

\subsubsection{Agent Random Envelope Variables}
The worst-case envelopes $\mathbf{E}_\text{worst}^{j,k}$ corresponding to a specific agent $j$ define random variable $\mathbf{E}_\text{worst}^{j}$ with each envelope getting assigned the probability difference covered by the contours $P(\mathbf{E}_\text{worst}^{j,k}) = p_k - p_{k-1}$.

\subsubsection{Combined Random Envelope Variables} Using a component-wise worst-case operation, denoted $\min$, we combine agent-specific random envelope variables into a single random envelope variable
\begin{equation}
	\mathbf{E}_\text{worst}^{comb.} = \min\Big[\mathbf{E}_\text{worst}^{1}, \mathbf{E}_\text{worst}^{2}, \ldots, \mathbf{E}_\text{worst}^{N}\Big]
\end{equation}
The random variable $\mathbf{E}_\text{worst}^{comb.}$ covers the true random envelope variable $\mathbf{E}$ in Eq.~\ref{eq:perception_envelope} in a safety consistent manner: It assigns equal probability to more or equal restrictive envelopes by using worst-case considerations over confidence levels and between agents assuming that the errors made due to sampling are neglectable.

\subsubsection{Resolving for Probabilistic Envelope}
Therefore, to fulfill Eq.~\ref{eq:perception_envelope}, we must calculate the cumulative distribution function of $\mathbf{E}_\text{worst}^{comb.}$ and resolve for $\mathbf{E_\text{percept}}$ such that it achieves a maximum cumulative probability $\beta_\text{risk}$.
Resolving the minimum operation by standard probability laws yields
\begin{equation} \label{eq:cdf_envelopes}
\begin{split}
\text{Pr}_{\delta_t \sim P_{\delta}}&(\mathbf{E}_\text{worst}^{comb.} < \mathbf{\mathbf{E}_\text{percept}}) = \\ &= 1 - \prod_{\forall j}P(\mathbf{E}_\text{worst}^{j} \geq \mathbf{E_\text{percept}}) = \\
 & = 1 - \prod_{\forall j}(1 - P(\mathbf{E}_\text{worst}^{j} < \mathbf{E}_\text{percept})) \overset{!}{\leq} \beta_\text{risk}
\end{split}
\end{equation}
We resolve this equation for $\mathbf{E_\text{percept}}$ component-wise.
For each acceleration component, the following steps are performed.
We iterate from most to least restrictive acceleration component of all envelopes $\mathbf{E}^{j,k}, \forall j, \forall k$, respectively and accumulate probabilities $P(\mathbf{E}_\text{worst}^{j,k})$ individually for each agent to estimate $P(\mathbf{E}_\text{worst}^{j} < \mathbf{\mathbf{E}_\text{percept}})$.
We check if the resulting combined cumulative probability $\text{Pr}_{\delta_t \sim P_{\delta}}(\mathbf{E}^{comb.} < \mathbf{E_\text{percept}})$ exceeds the allowed risk level $\beta_\text{risk}$. If yes, we use the last valid acceleration component within the envelope $\mathbf{\mathbf{E}_\text{percept}}$.

\subsection{Probabilistic Envelopes in the Context of Simplex Architecture} \label{sec:simplex_prob_envelope}
The resulting probabilistic safety envelope $\mathbf{E_\text{percept}}$ restricts the allowed accelerations of a nominal planning approach, e.g., a simple lane changing behavior. Given that the ego vehicle is witin a safe region at time $t$ the approach offers a guarantee on the probability of violating an envelope in the next time step, parameterizable with risk level $\beta_\text{risk}$. Yet, if the current state is already in violation of the envelope, these guarantee does not hold.

To deal with the current state violating the safety envelope, we extend the Simplex concept and use a probabilistic switching condition. Similar to the previous sections, we evaluate the violations at sampled states on the contours individually for each agent.
Worst-case considerations give a violation for contour $c_k$ if one of the sampled states is violated
\begin{equation}
v^{j,k} = \begin{cases} 1 \quad \exists p,q,r,\, f_\text{safety}(o_t^j+\delta_k(\phi_1^p, \phi_2^q, \phi_3^r))=1 \\
						0 \quad \text{else} \end{cases}
\end{equation}
Given the calculated violations for all agents and envelopes distributed according to $P(v^{j,k})=p_k-p_{k-1}$, we perform a probabilistic switch from nominal to safety behavior when the safety violation of a single agent $j$ is expected to exceed the risk level $\beta_\text{risk}$:
\begin{equation}
\exists j, \, \mathbb{E}_{v^{j,k}\sim P(\cdot)}\Big[v^{j,k}\Big] > \beta_\text{risk}
\end{equation}

%
%

\section{Experiments}
\newacronym{idm}{IDM}{intelligent driver model}
\begin{figure}[!t]
	\def\svgwidth{1.0\columnwidth}
	\centering
	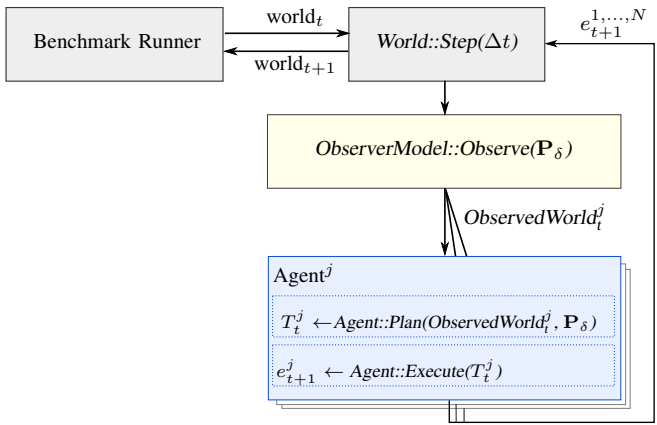
	\caption{Extended BARK architecture with an `ObserverModel' generating an $ObservedWorld^j_k$ for the j-th agent given the simulated measurement distribution $\mathbf{P}_\delta$. The $j$-th `Agent' plans a trajectory $T$ using the $ObservedWorld^j_t$ that is executed with an ideal trajectory controller.
	The agent has access to $\mathbf{P}_\delta$ to model an ideal perception.}
    \label{fig:results_observer_model}
    \vspace{-0.2cm}
\end{figure}
For evaluating the proposed probabilistic envelope, we use a highway scenario where the ego vehicle performs a lane change.
We show the insufficiency of conventional approaches, such as the \gls{rss}, in uncertain environments and benchmark it against our proposed probabilistic envelope method.

\subsection{Simulating Perception Uncertainty}
We extend the simulation platform BARK \cite{Bernhard2020} with an `ObserverModel' that adds noise to the j-th vehicle's state $\mathbf{s}^j_t$ to account for measurement and sensing uncertainties as shown in Figure~\ref{fig:results_observer_model}.
The observer model samples independent state deviations on top of the actual simulation state $\mathbf{s}^j_t$ in each simulation time-step $t$ resulting in an observed state $\mathbf{o}^j_t$ for each agent.
In this work, we use independent normal distributions to sample the state-values ($x,y,\theta, v$) of an observed state $\mathbf{o}^j_t$.
The measurement noise distribution $\mathbf{P}_\delta$ is used in the simulation to generate the samples $\mathbf{o}^j_t$ of the other vehicles that are passed to the ego vehicle's plan function to model an ideal estimate of measurement noise by a perception component ($\mathbf{\hat{P}}_\delta = \mathbf{P}_\delta$).
The parameters are shown in Table~\ref{tab:parameters} for the \emph{small covariance} and \emph{large covariance} case.
The ego vehicle's state is not subject to measurement noise and, therefore, fully observable.

\begin{table}[b]
	\scriptsize
	\vspace{-5mm}
	\centering
	\begin{tabular}{@{}llll@{}}
\toprule
                            & Name                                             & Value           & Unit                                                                                 \\ \midrule
\textbf{RSS Parameters}              & $\tau_\text{ego}$/$\tau_\text{other}$ & 0.2 / 1.0             & s                                                                                    \\
                            & $a_\text{min/max,lat}$                           & {[}-1.4, 1.4{]} &  m/$\text{s}^\text{\tiny 2}$ \\
                            & $a_\text{min/max,lon}$                           & {[}-8.0, 4.0{]} & m/$\text{s}^2$ \\
\textbf{Small Covariance}            & $\sigma_{\Delta x}$                              & 1.58             & m                                                                                    \\
                            & $\sigma_{\Delta y}$                              & 0.44             & m                                                                                    \\
                            & $\sigma_{\Delta v}$                             & 2.23             & m/s                                                                                  \\
                            & $\sigma_{\Delta \theta}$                         & 0.03           & rad                                                                                  \\
\textbf{Large Covariance}            & $\sigma_{\Delta x}$                              & 1.87             & m                                                                                    \\
                            & $\sigma_{\Delta y}$                              & 0.54             & m                                                                                    \\
                            & $\sigma_{\Delta v}$                             &    2.64           & m/s                                                                                  \\
                            & $\sigma_{\Delta \theta}$                         &    0.1         & rad                                                                                  \\
\textbf{Traffic Sampling Parameters} & Initial Long. Distances                          & {[}40, 50{]}    & m                                                                                    \\
                            & Initial Velocities $v$                           & {[}15, 20{]}    & m/s                                                                                  \\
 \bottomrule
\end{tabular}
	\caption{\gls{rss}, covariance and traffic parameters.} \label{tab:parameters}
\end{table}

\begin{figure}[!t]
	\centering
	\rotatebox{-90}{\input{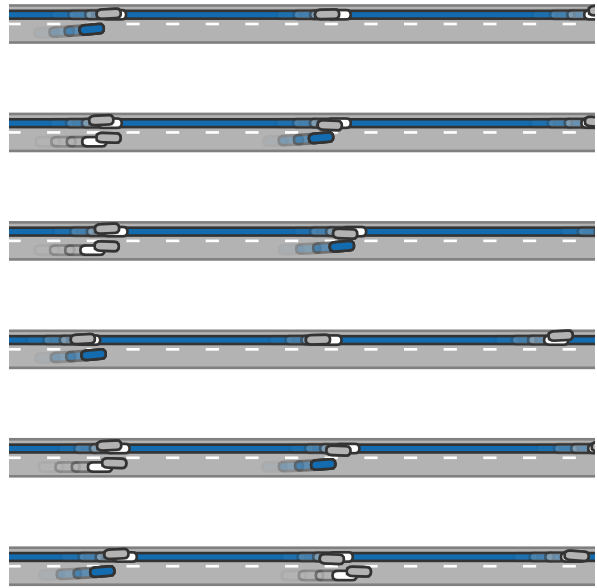}}
	\caption{The world states of the lane change scenario after the first four time steps ($t=0.8\text{s}$, $\Delta t=0.2 \text{s}$).
		The benchmarked agent (blue) intends to change to the left lane (goal definition is also depicted in blue).
		The past agent states are shown with increasing transparency, the white states are the ground-truth states, and the gray states are sampled by the `ObserverModel' at the current time step $t=0.8\text{s}$.}
	\label{fig:results_highway_scenario}
	\vspace{-0.15cm}
\end{figure}

\subsection{Scenario and Baseline Design}
We generate 100 scenarios with random ego and other vehicles' velocities ($[15.3\,\text{m/s}, 19.9 \,\text{m/s}]$) and initial longitudinal distances ($[40\,\text{m}, 50\,\text{m}]$).
The other vehicles are controlled by the \gls{idm} and the ego vehicle is controlled by one of the following approaches:
\begin{figure*}[!t]
	\centering
	\includegraphics[width=0.9\textwidth]{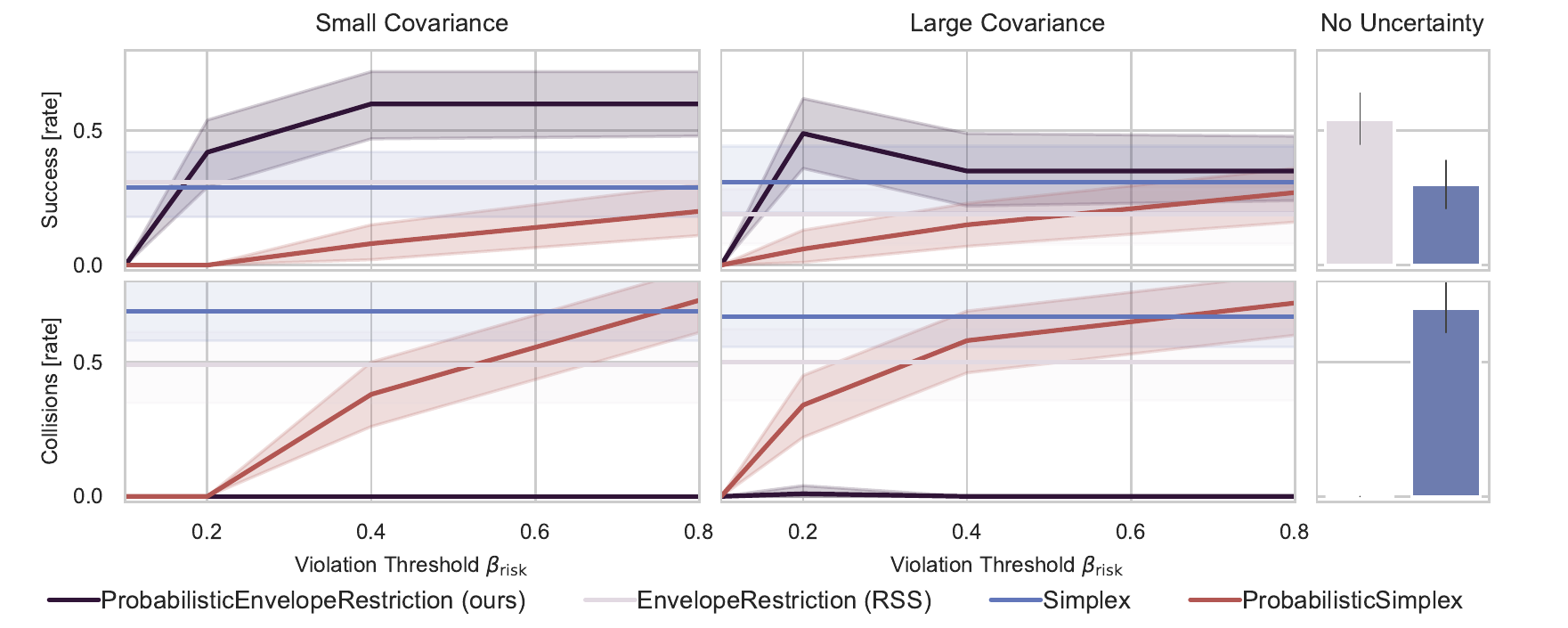}
	\caption{Performance results for the probabilistic envelope approach and baselines.}
	\label{fig:results}
	\vspace{-0.5cm}
\end{figure*}
\begin{itemize}
	\item \textbf{ProbablisticEnvelopeRestriction (ours):} The probabilistic envelope  $\mathbf{\mathbf{E}_\text{percept}}$ is calculated according to Sec.~\ref{sec:contour_lines} to \ref{sec:cdf_envelopes} restricting the accelerations of a lane changing controller.
	In case of detecting a violation of the current state according to conditions from Sec.~\ref{sec:simplex_prob_envelope}, the ego vehicle performs a safety maneuver with emergency braking and steering back to the right lane (without acceleration restrictions).
	\item \textbf{EnvelopeRestriction (\gls{rss}~\cite{shalev-shwartz_safe_2016}):} Similar to the previous model, but using the non-probabilistic envelope definition $\operatorname{env}(\cdot, \tau)$ to restrict the accelerations.
	In the case of violating the observed state $f_\text{safety}(\mathbf{o}_t)$, the safety maneuver is triggered.
	\item \textbf{Simplex:} No acceleration restrictions are applied to the controller.
	In the case of violating the observed state $f_\text{safety}(\mathbf{o}_t)$, the safety maneuver is triggered.
	\item \textbf{ProbabilisticSimplex:} This baseline randomly samples state deviations from the measurement noise distribution $\delta^j\sim \mathbf{P}_\delta(\cdot)$  to evaluate $f_\text{safety}(\mathbf{o}_t^j + \delta^j)$.
	If the sampled expectation over violations exceeds the risk level $\beta_\text{risk}$, the safety maneuver is triggered.  No acceleration restrictions are applied to the controller.
\end{itemize}

\subsection{Results}
All approaches, use an envelope definition $\operatorname{env}(\cdot, \tau)$ and violation detection $f_\text{safety}(\cdot)$ based on the open source \gls{rss} implementation\footnote{https://github.com/intel/ad-rss-lib} available in BARK.
The ego vehicle successfully completes a scenario if it reaches the (blue) goal region shown in Figure~\ref{fig:results_highway_scenario} within a pre-defined range of state limits (velocity, polygonal area, and angle) and 8 seconds simulation time.
Snapshots of the highway scenario depicting the actual and the sampled positions of the other vehicles are shown in Figure~\ref{fig:results_highway_scenario}.
Relevant parameters of all approaches are provided in Table~\ref{tab:parameters}.

For each baseline, we run the same 100 scenarios.
We analyze the success and collision rates and whether the maximum allowed scenario time is exceeded for increasing violation thresholds $\beta_\text{risk}$. Figure~\ref{fig:results} provides these rates for the case of no uncertainty, smaller and larger covariances of the measurement noise.
In the case of no uncertainty, \textbf{EnvelopeRestriction} and \textbf{ProbablisticEnvelopeRestriction} as well as \textbf{Simplex} and \textbf{ProbablisticSimplex} become conceptually equal.

Without uncertainty, the \textbf{EnvelopeRestriction} does not show any collisions and outperforms the \textbf{Simplex} approach significantly, which in contrast provoked a large number of collisions.
These collisions underline that purely reacting to violations of a safety specification using a safety maneuver is insufficient to achieve safety.

However, when simulating measurement uncertainty, the success rate of the \textbf{EnvelopeRestriction} drops and in 50\% of all scenarios collisions occur.
Since \textbf{EnvelopeRestriction} and \textbf{Simplex} do not depend on the violation threshold, the success and collision rates are drawn as horizontal lines in their respective color.
We observe that \textbf{ProbablisticSimplex} outperforms \textbf{EnvelopeRestriction} for lower violations thresholds  $\beta_\text{risk} \leq 0.4$ regarding the collision rate.
With increasing violation thresholds, the collision and success rate increase showing that already a simple sampling-based integration of uncertainty outperforms non-probabilistic approaches.
Yet, \textbf{ProbablisticSimplex} lacks a proper prevention of entering unsafe situations leading to a large number of collisions for increasing $\beta_\text{risk}$.
By restricting the acceleration in \textbf{ProbablisticEnvelopeRestriction}, success rates near 50\% are achieved without provoking any collisions in the \emph{small covariance} case.
For the \emph{larger covariance} case, collision occur at $\beta_\text{risk}=0.2$ while the overall success rate is lowered.

The results underline the meaningfulness of the problem formulation with $\beta_\text{risk}$ as parameter for quantitatively tuning allowed safety. The proposed \textbf{ProbablisticEnvelopeRestriction} naturally adapts the envelope to $\beta_\text{risk}$ and the amount of uncertainty.
Note that we deliberately chose high covariances for the evaluation to better capture differences between the baselines and our method.
Yet, we observe that collisions occur also with \textbf{ProbablisticEnvelopeRestriction} when $\beta_\text{risk}>0.1$. Also, for $\beta_\text{risk}>0$ collisions can occur with collision probability $P_\text{col.} \leq \beta_\text{risk}$ since $\beta_\text{risk}$ only defined how likely an envelope violation is. Since violating an envelope does not necessarily lead to a collision it holds $P_\text{col.} \leq \beta_\text{risk}$. The next section, gives an outline how to argue quantitative safety by specifying $\beta_\text{risk}$.

\section{Safety Argumentation Using Risk Levels}
We outline a safety argumentation for a highway pilot. We consider the severity as constant for this argumentation $\overline{severity}(highway)$. Since our work builds upon the RSS model, we assume that in absence of uncertainty, the RSS model guarantees no collision if the safety envelope is not violated. Hence, the source of collision is assumed to be the perception uncertainty causing envelope violations in the ego vehicle.
Given this context, the top level assurance goal is:


\textbf{Goal G}: \textsl{"The highway pilot shall achieve a responsible-sensitive collision risk $\rho_\text{col}(\text{highway})$"}. Responsible-sensitive denotes the fact that $\rho_\text{col}(\text{highway})$ controls the risk of being responsible for a collision when other vehicles \emph{do not violate} their safety envelopes. The collision risk is e.g. set to $\rho_\text{col}(\text{highway})=10^{-7}$. Next, we derive two sub-goals using

\textbf{Strategy S:} \textsl{"Separation of collision risk based on maneuver classes and maneuver frequency"}. This strategy relies on the frequency of maneuvers, e.g. for merging situations $P(\text{merge})$ being standardized using data, similar to the definition of exposure rates defined in ISO26262.
Note that the number of decomposed sub-goals depends on the number of maneuvers taken into consideration. We argue for the merge maneuver with others having a similar methodology.

\textbf{Goal G1}: \textsl{"The highway pilot shall achieve a collision risk $\rho_\text{col}(\text{following}) \overset{!}{=} 0$ when following another car"}.
This subgoal is assured by \text choosing $\beta_\text{risk} = 0$ for the case of when following another car. The conservativeness of an envelope obtained from $\beta_\text{risk} = 0$ is acceptable in a pure lane following situation.

\textbf{Evidence E1} \textsl{"With $\beta_\text{risk} = 0$ it holds $\rho_\text{col}(\text{following}) = 0$ due to $P_\text{col} \leq \beta_\text{risk}$}

\textbf{Goal G2}: \textsl{"The highway pilot shall achieve a collision risk $\rho_\text{col}(\text{merge}) \overset{!}{<} \rho_\text{col}(\text{highway}) / P(\text{merge})$ during merging"}. The collision risk $\rho_\text{col}(\text{merge}) = \overline{severity}(highway)\cdot P_\text{col}(merge)$ is a combination of severity and collision probability $P_\text{col}(merge)$ during merging. This gives
\begin{equation}
\begin{split}
&\overline{severity}(highway)\cdot P_\text{col}(merge) \leq \frac{\rho_\text{col}(\text{highway})}{P(\text{merge})} \\
&\implies P_\text{col}(merge) \leq \frac{ \rho_\text{col}(\text{highway}) }{P(\text{merge})\overline{severity}(highway)}
\end{split}
\end{equation}
Note that we could set $\beta_\text{risk} = P_\text{col}(merge)$ to fulfill this requirement. Yet, for low levels of $\beta_\text{risk}$ very restrictive envelopes as seen in our experiment arise and solving becomes numerically unstable. Instead, we achieve this goal using two evidences

\textbf{Evidence 2.1:} \textsl{"Violating the true safety envelope during merging does lead with probability less than $P_{\text{violate}\rightarrow\text{col}}(\text{merge})$ to a collision"} Studies have shown that humans frequently violate envelopes without provoking collisions \cite{esterle_formalizing_2020}. These findings must be standardized depending on situation and scenarios. 

\textbf{Evidence 2.1:} \textsl{"The highway pilot does violates the true safety envelope with probability less than $\beta_\text{risk} =  P_\text{col}(merge) /P_{\text{violate}\rightarrow\text{col}}$"}

Presented  argumentation will be refined as future work. It reveals that the presence of uncertainty requires an extended form of safety argumentation compared to "yes/no" argumentation schema of non-probabilistic envelopes such as RSS.

\section{Conclusion}
Approaches like the \gls{rss} offer safety guarantees when there are only prediction uncertainties present and if all vehicles adhere to it (remain in the defined safety envelope). However, practical applications require integration of sensing and localization uncertainties, and it is essential to consider these in a sense, plan and act cycle. In this work we developed a risk-based probabilistic safety envelope approach for calculating safety envelopes under perception uncertainty. The risk threshold limits the probability that the true safety envelope in a fully observable environment is larger than the applied safety envelope. We showed that our approach outperforms conventional safety envelopes such as RSS, that do not consider uncertainties. Furthermore, it provides quantifiable probabilistic safety, with risk lower than a specified threshold. Finally, we outlined a safety argumentation for highway merging scenarios, which we plan to refine in future work.

\section*{Acknowledgement}
This research was partly funded by the Providentia++ project
and the ``Künstliche Intelligenz Europäisch Zertifizieren
unter Industrie 4.0" (KIEZ 4.0) project.

\appendices


\AtNextBibliography{\footnotesize}
\printbibliography

\end{document}